\documentclass[%
 aip,
 amsmath,amssymb,
preprint,%
 reprint,%
,groupedaddress]{revtex4-1}

\usepackage{graphicx}
\usepackage{dcolumn}
\usepackage{bm}

\usepackage[utf8]{inputenc}
\usepackage[T1]{fontenc}
\usepackage{mathptmx}
\usepackage{etoolbox}
\usepackage[]{fancyhdr}

\makeatletter
\def\@email#1#2{%
 \endgroup
 \patchcmd{\titleblock@produce}
  {\frontmatter@RRAPformat}
  {\frontmatter@RRAPformat{\produce@RRAP{*#1\href{mailto:#2}{#2}}}\frontmatter@RRAPformat}
  {}{}
}%
\makeatother

\draft 

\begin{document}
\fancyhead[C]{The following article has been submitted to Applied Physics Letters. After it is published, it will be found at https://aip.scitation.org/journal/apl}
\title{Probing itinerant carrier dynamics at the diamond surface using single nitrogen vacancy centers} 



\author{Marjana Mahdia}
\affiliation{Department of Electrical and Computer Engineering, Princeton University, Princeton, New Jersey 08544, USA}
\author{James Allred}
\affiliation{Department of Electrical and Computer Engineering, Princeton University, Princeton, New Jersey 08544, USA}
\author{Zhiyang Yuan}
\affiliation{Department of Electrical and Computer Engineering, Princeton University, Princeton, New Jersey 08544, USA}
\author{Jared Rovny}
\affiliation{Department of Electrical and Computer Engineering, Princeton University, Princeton, New Jersey 08544, USA}
\author{Nathalie P. de Leon}%
\email[]{npdeleon@princeton.edu}
\affiliation{Department of Electrical and Computer Engineering, Princeton University, Princeton, New Jersey 08544, USA}


\date{\today}

\begin{abstract}
Color centers in diamond are widely explored for applications in quantum sensing, computing, and networking. Their optical, spin, and charge properties have been extensively studied, while their interactions with itinerant carriers are relatively unexplored. Here we show that NV centers situated within 10 nm of the diamond surface can be converted to the neutral charge state via hole capture. By measuring the hole capture rate, we extract the capture cross section, which is suppressed by proximity to the diamond surface. The distance dependence is consistent with a carrier diffusion model, indicating that the itinerant carrier lifetime can be long, even at the diamond surface. Measuring dynamics of near-surface NV centers offers a new tool for characterizing the diamond surface and investigating charge transport in diamond devices.
\end{abstract}


\maketitle 
\thispagestyle{fancy} 

Color centers have been widely studied for their applications in quantum sensing, quantum networks, and quantum information processing \cite{childress2013diamond, hong2013nanoscale, schirhagl2014nitrogen}. Nitrogen vacancy (NV) centers in diamond in particular are an attractive platform because they exhibit long spin coherence times at room temperature and they allow for off-resonant optical detection and initialization of spin states \cite{childress2013diamond, Bar}. Charge state stability and control of NV centers is of particular interest for applications in nanoscale sensing\cite{staudacher2013nuclear, mamin2013nanoscale}, superresolution microscopy\cite{chen2015subdiffraction} and long-term data storage\cite{dhomkar2016long}. Recent experiments have focused on selectively preparing \cite{kurtsiefer2000stable, rondin2010surface, cui2013increased, schreyvogel2015active} and reading out \cite{neumann2010single, robledo2011high, shields2015efficient} particular charge states, as well as studying the impact of charge dynamics on optically detected magnetic resonance \cite{yuan, dolev}. However, the interactions between itinerant carriers and color centers is less well explored, and can strongly impact the color center charge state, ionization dynamics, and spin readout. Such interactions could also be harnessed for new functionality such as electrically detected magnetic resonance, \cite{hrubesch2017efficient, siyushev2019photoelectrical} and stabilizing non-equilibrium charge distributions \cite{dhomkar2018demand, zhang2022neutral}. 

Recent work has focused on using itinerant carriers to manipulate color centers in the diamond bulk. For example, it was shown that the optically dark state of a silicon vacancy (SiV) center is SiV\textsuperscript{2-} through charge state readout of NV centers and SiV centers combined with remote optical pumping \cite{gardill2021probing}. In another example, holes generated by one NV center were captured by another NV center, converting the latter NV center to the neutral charge state \cite{lozovoi}. Both studies examined NV centers far (>10 $\mu$m) from the surface, at which distances surface effects are negligible.

\begin{figure}
\centering
\includegraphics{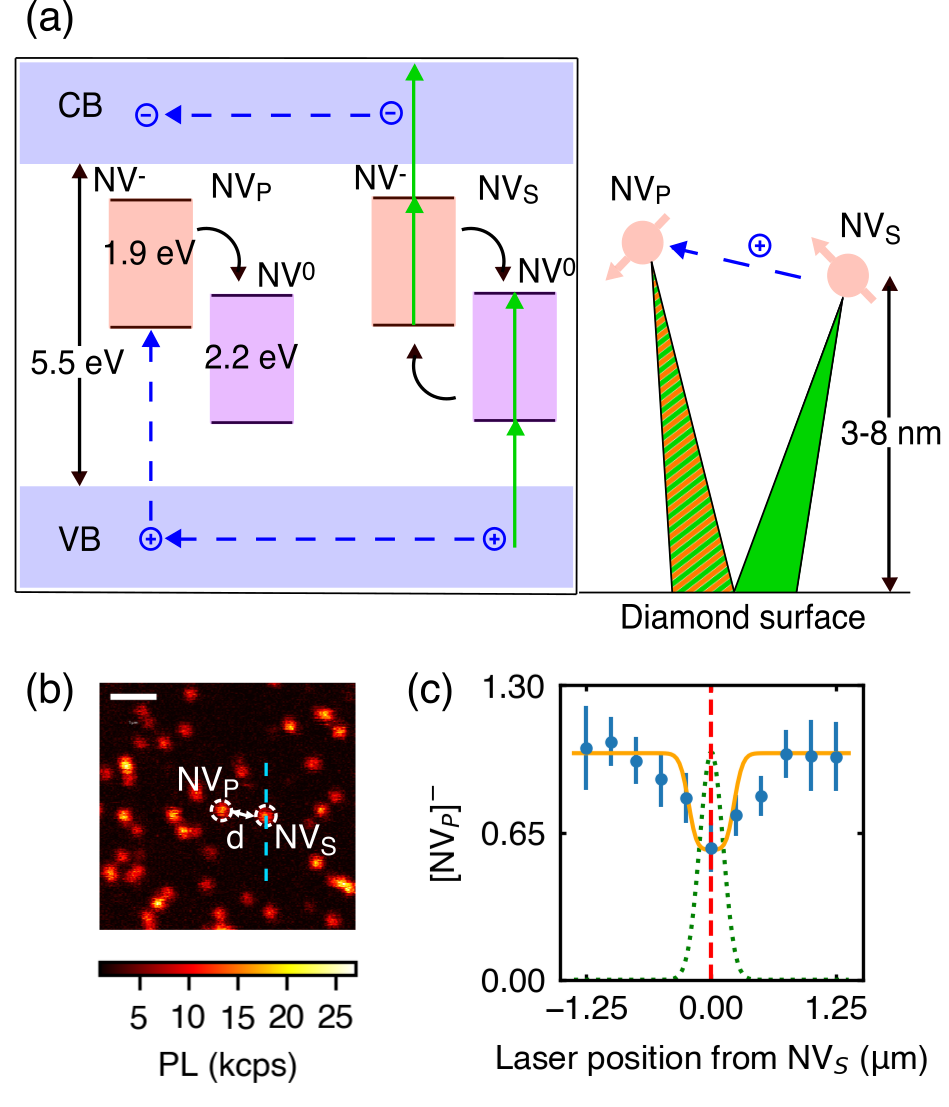}
\caption{(a) Energy level diagram showing charge state conversion processes between NV\textsubscript{S} and NV\textsubscript{P}. Solid vertical arrows indicate photoionization processes, dashed arrows indicate itinerant carrier transport and capture, and curved arrows indicate charge state conversion. (b) Scanning confocal microscope image of two NV centers. The scale bar is 1 $\mu$m. (c) The NV\textsuperscript{-} charge state population of NV\textsubscript{P} as a second laser is scanned along the line cut shown in (b) with the NV\textsubscript{S} position indicated with the red dashed line. The orange solid line is a fit considering the dependence of [NV\textsubscript{P}]\textsuperscript{-} on the NV\textsubscript{S} excitation laser power. The laser spot size is indicated for reference (green dotted line).}
\label{fig:figure1}
\end{figure}

Here we study charge dynamics of shallow NV centers <10 nm from the surface and their interaction with itinerant carriers. Shallow defects are essential for high sensitivity quantum sensing, and understanding the NV charge state and itinerant carrier dynamics near the surface is critical for developing shallow NV centers as a quantum platform. We demonstrate that the charge state of a shallow probe NV center (NV\textsubscript{P}) can be controlled by free carriers generated by excitation of another remote shallow source NV center (NV\textsubscript{S}) up to $\sim$ 7 $\mu$m away (Fig. \ref{fig:figure1}(a, b)). Specifically, the charge state of NV\textsubscript{P} converts from negative to neutral as a second 532 nm excitation laser is scanned over NV\textsubscript{S} (Fig. \ref{fig:figure1}(c)). Continuous optical ionization and recombination of NV\textsubscript{S} generates a constant flow of holes and electrons in the valence and conduction bands, respectively. These itinerant carriers can diffuse and subsequently be captured by NV\textsubscript{P}. The net conversion to the neutral charge state implies that hole capture is much more efficient than electron capture, leading to a net change in the steady-state charge. This phenomenon has been observed in some previous works \cite{gardill2021probing, lozovoi}.

\begin{figure}

\centering
\includegraphics{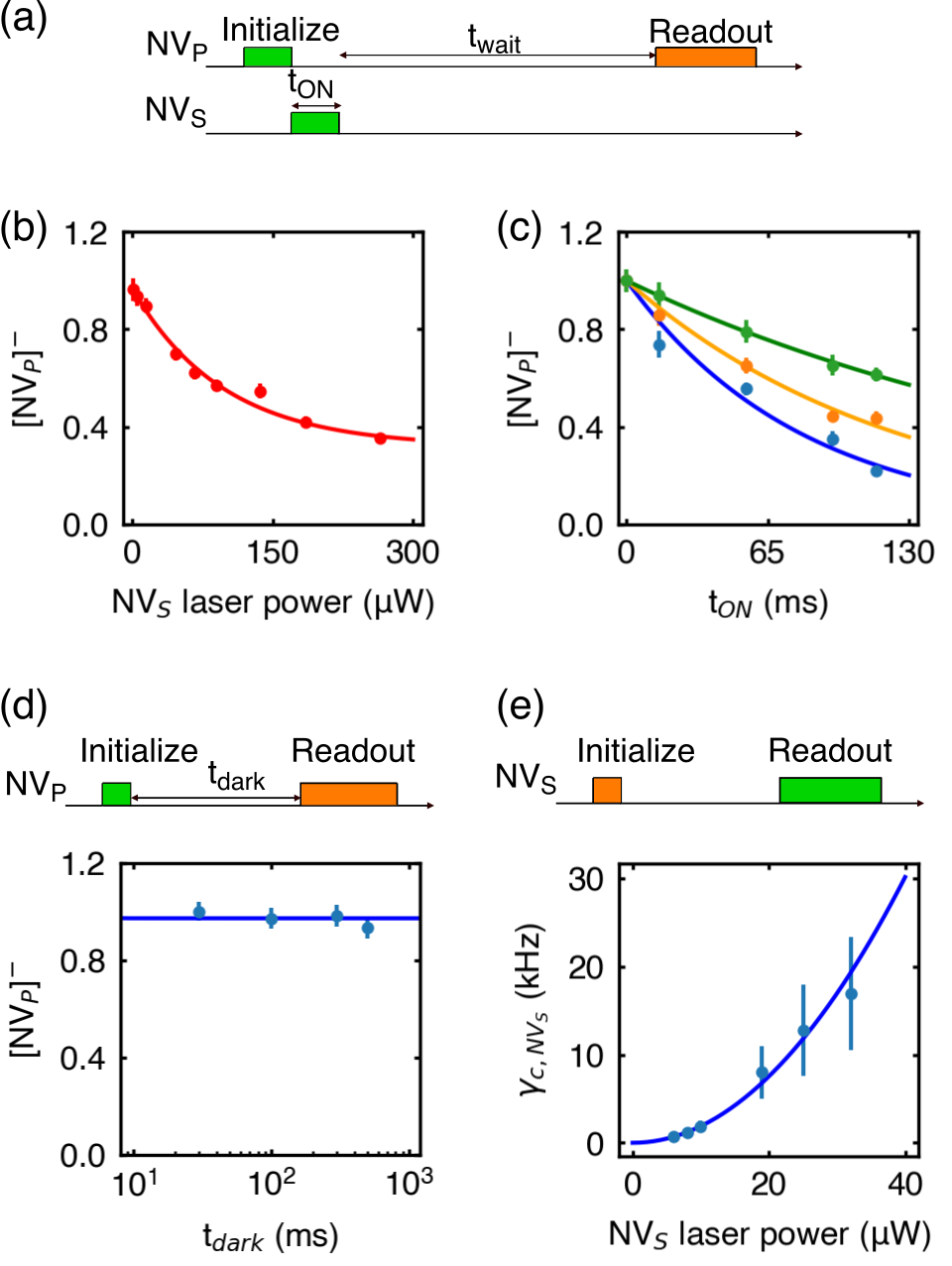}
\caption{(a) Pulse sequence used to detect the change of NV\textsubscript{P} charge state due to optical excitation of NV\textsubscript{S}. A 532 nm laser pulse (94 $\mu$W, 5ms) initializes NV\textsubscript{P}. Another 532 nm pulse (variable power) continuously drives ionization and recombination processes of NV\textsubscript{S} for t\textsubscript{ON}. A 591 nm laser pulse (2 $\mu$W, 150 ms) is used to readout the charge state of NV\textsubscript{P}. A wait time, t\textsubscript{wait} = 15 ms is introduced to avoid background phosphorescence from optical components. (b) Dependence of [NV\textsubscript{P}]\textsuperscript{-} on NV\textsubscript{S} excitation laser power (t\textsubscript{ON}=15 ms). The solid line indicates an exponential fit. (c) [NV\textsubscript{P}]\textsuperscript{-} as a function of t\textsubscript{ON} for different NV\textsubscript{S} excitation laser powers (green: 138 $\mu$W, orange: 168 $\mu$W, blue: 216 $\mu$W), $d$ = 2.6 $\mu$m. Solid lines indicate exponential fits. (d) The stability of the [NV\textsubscript{P}]\textsuperscript{-} in the dark. No change is detected out to 500 ms. (e) Hole generation rate ($\gamma_c$) versus green (readout) laser power for NV\textsubscript{S}. The fit curve is quadratic, indicating a two photon process.}
\label{fig:figure2}
\end{figure}

\begin{figure}

\centering
\includegraphics{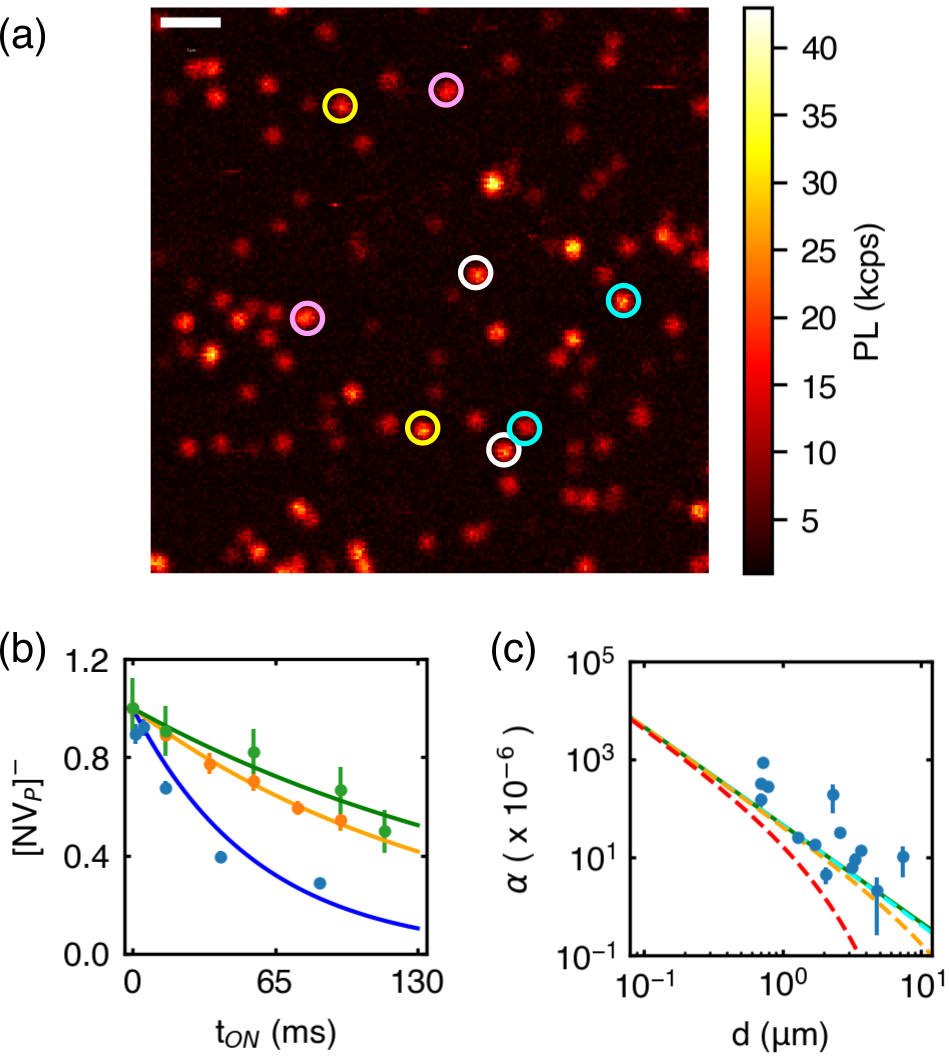}
\caption{(a) Confocal scan showing a subset of the NV pairs investigated for calculating $\sigma\textsubscript{h}$ and $L$\textsubscript{h}. Pairs are indicated by different colors. The scale bar is 1 $\mu$m. (b) Time-dependent decay of [NV\textsubscript{P}]\textsuperscript{-} varies with $d$ (green: 1.71 $\mu$m, orange: 1.29 $\mu$m, blue: 0.70 $\mu$m). NV\textsubscript{S} is excited with a 90 $\mu$W 532 nm laser. Solid lines indicate exponential fits. (c) The ratio of the hole capture rate to the hole generation rate, $\alpha$, for the 14 different NV pairs under study. The distance dependence is consistent with a diffusion model with an infinite diffusion length (green solid line). For comparison, three dashed lines with finite $L$\textsubscript{h} (cyan: 100 $\mu$m, orange: 10 $\mu$m, red: 1 $\mu$m) are also included.}

\label{fig:figure3}
\end{figure}

NV centers are individually interrogated using a dual channel, multicolor confocal microscope (see methods). In the experiment, two optically resolvable NV centers are chosen within the scanning field of view, NV\textsubscript{P} and NV\textsubscript{S}. One optical channel monitors the charge state of NV\textsubscript{P} after initialization into NV\textsuperscript{-} with roughly 70\% probability using a high power green pulse (94 $\mu$W, 5 ms), followed by charge state readout with a low power orange pulse (2 $\mu$W, 150 ms). In between these pulses, the second optical channel is used to pump NV\textsubscript{S} for a time t\textsubscript{ON} with a green laser pulse of variable power, cycling its charge state between NV\textsuperscript{-} and NV\textsuperscript{0} (Fig. \ref{fig:figure2}(a)). This cycling generates free carriers that diffuse away and can be subsequently captured by nearby NV centers. The charge state population is measured by obtaining fluorescence with orange excitation since 591 nm is situated in between the zero phonon lines (ZPL) of NV\textsuperscript{-} (637 nm) and NV\textsuperscript{0} (575 nm). A time window shorter than the decay time obtained from the fluorescence trace is defined, over which a histogram of photon counts is drawn. The area under the double Poisson shaped histogram is used to determine the charge state population of NV\textsubscript{P} (see Supplementary Material for details).

We interrogate the charge capture kinetics by measuring the charge state of NV\textsubscript{P} while varying the excitation power at NV\textsubscript{S}. As the excitation power increases, the population of the negative charge state ([NV\textsubscript{P}]\textsuperscript{-}) decreases (Fig. \ref{fig:figure2}(b)). Moreover, [NV\textsubscript{P}]\textsuperscript{-} decays exponentially with the duration of the NV\textsubscript{S} excitation pulse, and the decay time constant decreases with increasing excitation power (Fig. \ref{fig:figure2}(c)). Without any excitation at NV\textsubscript{S}, there is no change in [NV\textsubscript{P}]\textsuperscript{-} over time (Fig. \ref{fig:figure2}(d)). The net decay of [NV\textsubscript{P}]\textsuperscript{-} during NV\textsubscript{S} illumination indicates that hole capture in the negative charge state dominates over electron capture in both the negative and neutral charge states.
 
We model the hole capture rate, $\gamma_h$ as:
\begin{equation}
\begin{split}
\gamma_{h} & = \rho_hc_h + \gamma_d, \\
& = \frac{2\gamma_ce^{-d/L_h}}{4\pi d^2\nu_h}\sigma_h\nu_h + \gamma_d, \\
& = \frac{\sigma_h}{2\pi d^2}\gamma_ce^{-d/L_h} + \gamma_d.
\end{split}
\label{eqn:equation1}
\end{equation}
where $\rho_h$ is the hole density, $c_h$ is the capture coefficient, $\gamma_c$ is the hole generation rate at NV\textsubscript{S}, $L_h$ is an effective hole diffusion length that arises from the free carrier lifetime, $d$ is the distance between NV\textsubscript{S} and NV\textsubscript{P}, $\sigma_h$ is the hole capture cross-section at NV\textsubscript{P}, $\nu_h$ is the hole velocity, and $\gamma_d$ is the dark ionization rate, which is constant. The factor of two in the numerator of equation (\ref{eqn:equation1}) arises from a geometrical factor--we assume that carriers can reflect from the surface, and the NV centers are much closer to the surface than they are to each other. To calculate $\gamma_c$ (Fig. \ref{fig:figure2}(e)), we first fit the fluorescence of NV\textsubscript{S} under 532 nm illumination (after initialization with 591 nm laser) to an exponential fit to extract the total charge conversion rate, $\gamma_{total}$ = $\gamma_i$ + $\gamma_r$ of NV\textsubscript{S}, where $\gamma_i$ and $\gamma_r$ are ionization and recombination rates respectively\cite{yuan} (see Supplementary Material for details). We measure the NV\textsuperscript{-} population of several NV centers in our sample at steady state for several powers of 532 nm initialization. All of the measured NV centers show NV\textsuperscript{-} population in the range of [55\%, 70\%], which along with $\gamma_{total}$ is used to calculate $\gamma_r$, and subsequently $\gamma_i$. In steady state, since the time between subsequent holes is the total time it takes for a hole and an electron to be generated, we have $\gamma_c = (1/\gamma_i+1/\gamma_r)^{-1}$. 

We measured 17 pairs of NV centers in total and observed hole capture in 14 of the pairs (Fig. \ref{fig:figure3}(a)). The hole capture rate varies among pairs of NV centers, and is generally slower for NV pairs with larger spacing $d$ (Fig. \ref{fig:figure3}(b)). This distance dependence could arise from the area scaling of carrier diffusion or from a finite carrier lifetime. By rearranging equation (\ref{eqn:equation1}) we can define a parameter $\alpha$ to investigate if the hole carrier lifetime is an important factor, where
\begin{equation}
\alpha = \frac{\gamma_h-\gamma_d}{\gamma_c} = \frac{\sigma_h e^{-d/L_h}}{2\pi d^2}.
\label{eqn:equation2}
\end{equation}

We assume $\gamma_d$ = 0 because the run time of the experiment (< 600 ms) is shorter than the dark lifetime of NV\textsubscript{P} (Fig. \ref{fig:figure2}(c), see Supplementary Material for details). The calculated $\alpha$ for each NV pair is plotted versus inter-NV distance in Fig. \ref{fig:figure3}(c). The distance dependence is consistent with a 1/$d^2$ scaling (see Supplementary Material for details). We therefore conclude that the effective ionization of NV\textsubscript{P} due to NV\textsubscript{S} is not limited by the diffusion length of holes. 

From the fit in Fig. \ref{fig:figure3}(c), we extract the capture cross-section, $\sigma\textsubscript{h}$ = 2.89$\times$10\textsuperscript{-4} $\pm$ 0.54$\times$10\textsuperscript{-4} $\mu$m$^{2}$. The large value of $\sigma\textsubscript{h}$ likely arises from the Coulomb attraction between the negatively charged NV center and the hole, resulting in Rydberg-like states \cite{zhang2020optically}. We note that although the cross section is large, this value is an order of magnitude smaller than previously reported for deep NV centers \cite{lozovoi}. The surface-related suppression of hole capture could arise from finite hole lifetime due to surface traps or reduction in the effective cross section because of geometric overlap with the surface. We rule out the former reason based on the distance dependence shown in Fig. \ref{fig:figure3}(c).  

We have demonstrated generation and capture of free carriers between two shallow NV centers that are $<$10 $\mu$m apart from one another. We have shown that the hole capture cross section is smaller than prior measurements of bulk NV centers, but that the observed carrier capture rate is not limited by the carrier lifetime. The hole diffusion length and hole capture cross section can be utilized as sensitive probes of charge transport in diamond devices. The technique demonstrated here can be easily extended to stabilize particular charge states of defects through photo-doping with distant donors, rather than bulk doping, as we have recently demonstrated for SiV\textsuperscript{0} centers \cite{zhang2022neutral}. A natural next step would be to deploy photodoping to stabilize new color centers such as GeV\textsuperscript{0}, SnV\textsuperscript{0} and PbV\textsuperscript{0}.

\section*{METHODS}

Diamond samples are prepared using the same process detailed in the previous work\cite{peace}. Diamond substrates from Element Six (E6) are laser cut and scaife polished to a (100) face within a 3 degree specification. They are then reactive ion etched under Ar/Cl\textsubscript{2} and then under O\textsubscript{2}. This etching damage is removed by a 1200$^\circ$C vacuum anneal. The resulting amorphous carbon layer is removed by refluxing a 1:1:1 mixture of perchloric, nitric, and sulfuric acids for at least two hours (triacid clean). The samples are then implanted with \textsuperscript{15}N at 3 keV at a dose of 0.5$\times$10\textsuperscript{9} cm$^{-2}$. Following ion implantation, samples are triacid cleaned and then vacuum annealed at 800$^\circ$C to form nitrogen-vacancy complexes. Following another triacid clean, the samples are then oxygen annealed at 460$^\circ$C to form a well-ordered oxygen-terminated surface.

To investigate the charge state modification, NV centers are interrogated using multicolor confocal microscope with two excitation pathways. These excitation pathways each pass through two galvanometers (model: Thorlabs GVS012) before being combined into the objective. Photons collected from the sample are counted through two avalanche photodiodes (model: Excelitas SPCM-AQRH-44-FC). This allows for simultaneous optical control and measurement of two distinct NV centers. One excitation path contains only 532 nm laser (laser on NV\textsubscript{S} in Fig. \ref{fig:figure2}(a) pulse sequence) from a laser source (model: Coherent Sapphire 532-300 LP), and is used to cycle the charge state of NV\textsubscript{S}. The other excitation path (laser on NV\textsubscript{P} in Fig. \ref{fig:figure2}(a) pulse sequence) includes light from a continuum source (model: NKT SuperK, repetition rate 78 MHz, pulse width 5 ps), bandpass-filtered at 591 nm, in addition to 532 nm light from another branch of the same green laser source mentioned before. This path is used to interrogate the charge state of NV\textsubscript{P}. This setup is similar to the setup used in previous work\cite{yuan}, the primary difference being the path interrogating NV\textsubscript{S}.

Laser light sources are coupled into acousto-optic modulators, allowing for pulses with sub 100 ns rise and fall times. A Pulse Blaster (model: SpinCore ESR-PRO500 with a timing resolution of 2 ns) is used to control the acousto-optic modulators and the timing of readout. 

The photon count rate depends sensitively on the charge state of the NV center. As a result, a histogram of photon counts per trial reveals two peaks corresponding to the two charge states. Fitting these peaks to a double Poisson distributions then reveals the $\text{NV}^-$ charge state probability.

\label{sec:methods}

\section*{SUPPLEMENTARY MATERIAL}
See the Supplementary Material for details about determination of NV\textsuperscript{-} population, calculation of carrier generation rate, diagnosis of dark ionization rate, verification of long diffusion length, and hole capture rates for concerned NV centers.

\begin{acknowledgments}
We would like to thank Carlos Meriles and Artur Lozovoi for helpful discussions. This work was primarily supported by the US Department of Energy, Office of Science, Office of Basic Energy Sciences, under Award No. DE-SC0018978. Diamond surface preparation was supported by the NSF under the CAREER program (grant DMR1752047). J.A. acknowledges the NSF Graduate Research Fellowship Program for support. J.R. acknowledges the Princeton Quantum Initiative Postdoctoral Fellowship for support.
\end{acknowledgments}

\section*{Author Declarations}
\subsection*{Conflict of Interest}
The authors have no conflicts to disclose.
\subsection*{Author Contributions} M.M., J.A., J.R., and N.P.d.L. conceptualized the project, designed experiments, analyzed the data, and wrote the manuscript. M.M. and J.A. carried out experiments, Z.Y. and J.R. helped optimize the experimental apparatus and charge state readout. All authors contributed to writing and editing the manuscript.

\section*{Data Availability}
The data that support the findings of this study are available from the corresponding author upon reasonable request.


\section*{References}
\bibliography{References.bib}

\end{document}


\maketitle

\section*{A. NV\textsuperscript{-} population determination}
\begin{figure}[h]
\centering
\includegraphics[width=1\linewidth]{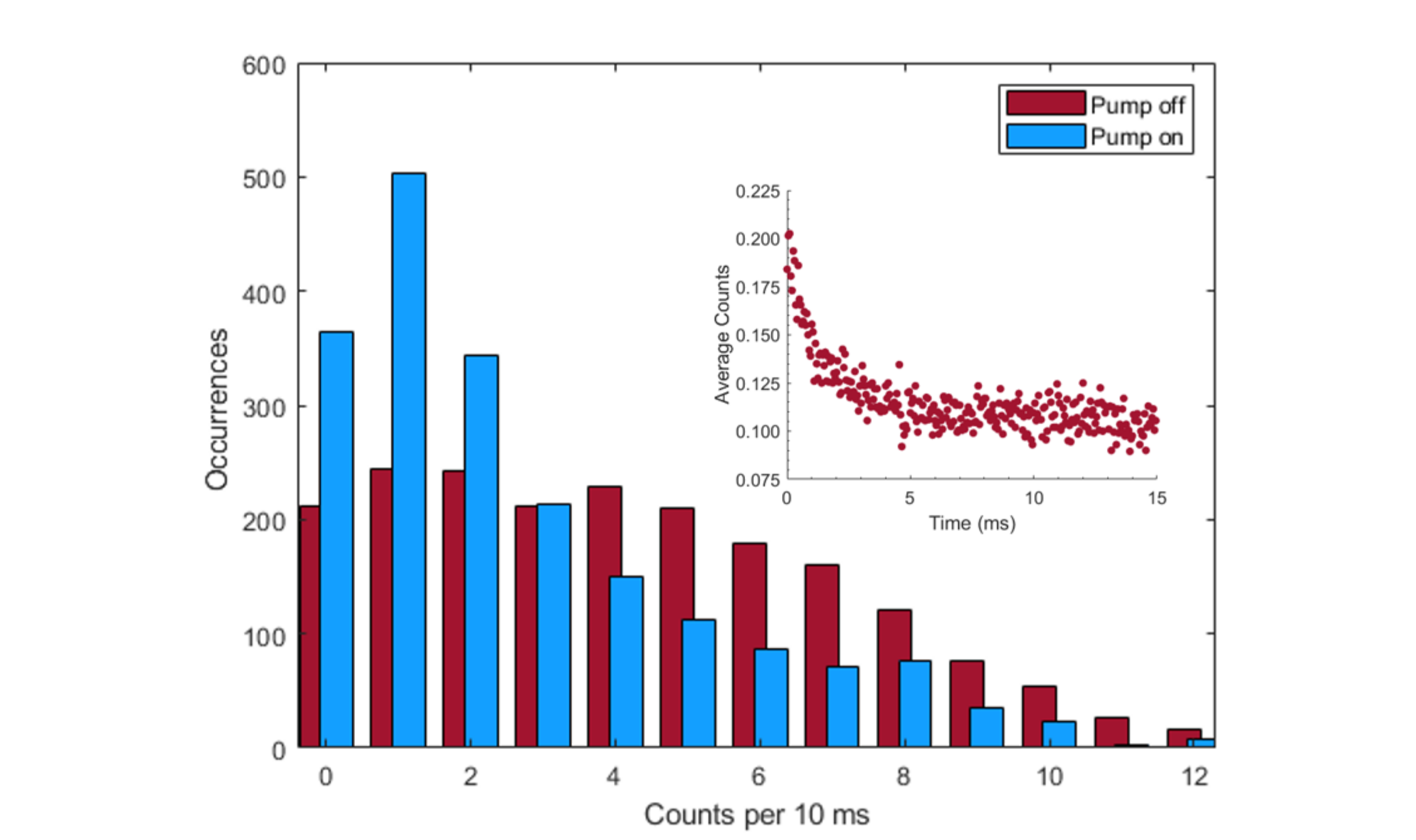}

\caption{Determination of [NV\textsubscript{P}]\textsuperscript{-} from double Poisson fit. Blue (red) diagram corresponds to laser on NV\textsubscript{S} being ON (OFF). 
\label{fig:si_fig1}}
\end{figure}

To determine the NV\textsuperscript{-} population of NV\textsubscript{P} ([NV\textsubscript{P}]\textsuperscript{-}), the pulse sequence in Fig. 2a (main text) is used. After initializing NV\textsubscript{P} with a 532 nm laser pulse, another 532 nm laser pulse is used to pump NV\textsubscript{S}. During the readout period of NV\textsubscript{P} with the orange laser pulse, we see an exponential decay in the fluorescence (Fig. \ref{fig:si_fig1} inset). The decay time is determined from an exponential fit of the fluorescence, which defines the length of an integration window. A histogram of photon counts during this window in many experiment iterations is plotted (Fig. \ref{fig:si_fig1}). The histogram can be fitted with a double Poisson function. The area under the Poisson fit of higher (lower) average photon counts gives us the NV\textsuperscript{-} (NV\textsuperscript{0}) population probability of NV\textsubscript{P}.

\pagebreak
\section*{B. Carrier generation rate calculation}
\begin{figure}[h]
\centering 
\includegraphics[width=1\linewidth]{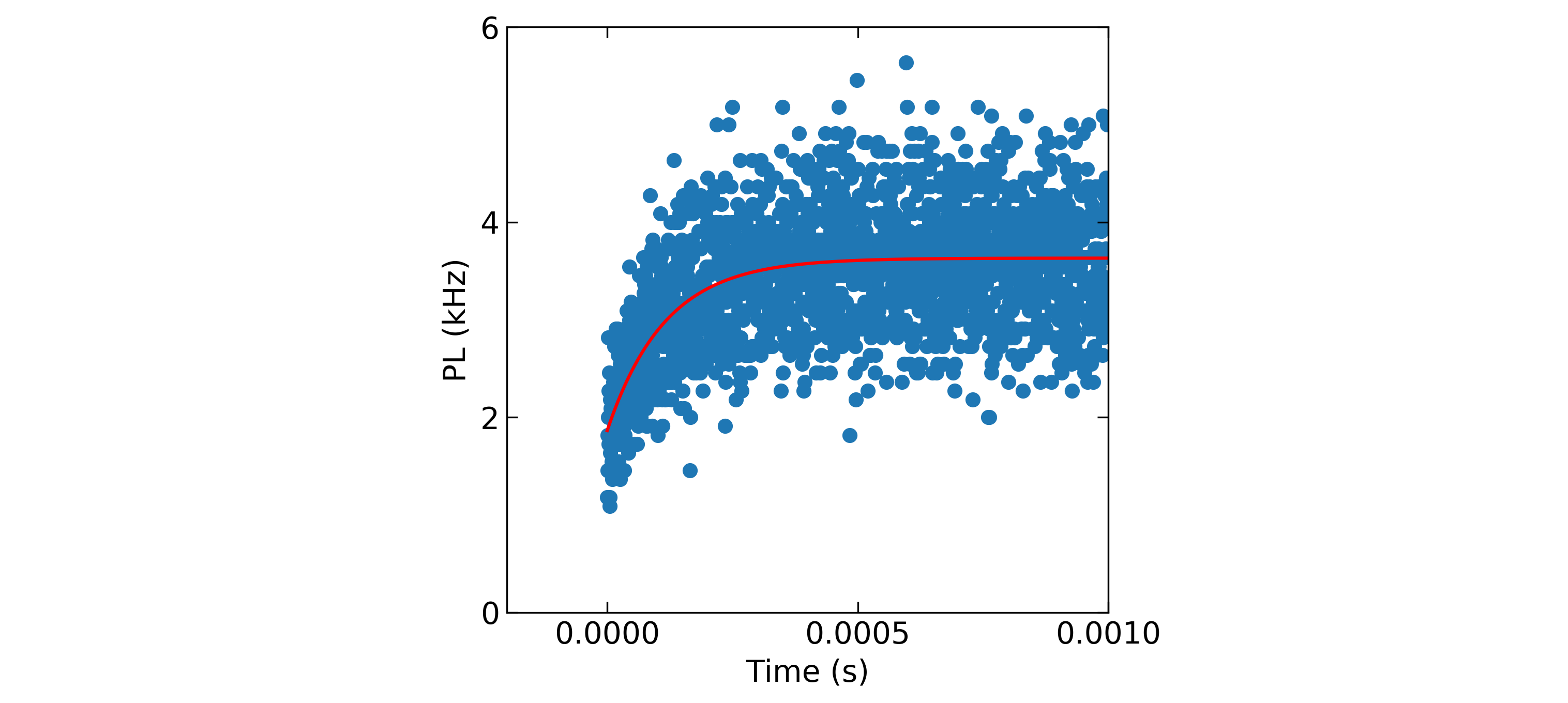}
\caption{Calculation of carrier generation rate of NV\textsubscript{S}. Red line is an exponentially rising fit of the fluorescence. 
\label{fig:si_fig2}}
\end{figure}

Using the pulse sequence shown in Fig. 2e (main text), the carrier generation rate of NV\textsubscript{S} is determined. NV\textsubscript{S} is at first initialized into mostly NV\textsuperscript{0} charge state with an orange laser, and then a green laser is applied for readout. During the readout, we see an exponential rise in the fluorescence, which can be fitted with an exponential fit (red solid line in Fig. \ref{fig:si_fig2}). The exponential fit provides us with a rise time, from which we calculate $\gamma_{total}$. We have $\gamma_{total} = \gamma_{i} + \gamma_{r}$, where $\gamma_{i}$, $\gamma_{r}$, and $\gamma_{total}$ are the ionization rate, recombination rate, and total rate for a particular excitation wavelength and power, respectively. We calculate $\gamma_r$ from the equation: $\gamma_r = [NV]\textsuperscript{-} \gamma_{total}/100$, where [NV]\textsuperscript{-} is the steady state NV\textsuperscript{-} population of NV\textsubscript{S} under green excitation. For different powers of green excitation, we measure [NV]\textsuperscript{-} for several NV centers, which falls in the range of [55\%, 70\%]. We use this range to calculate $\gamma_r$, and subsequently $\gamma_i$. We then calculate the hole (or, electron) carrier generation rate, $\gamma_{c} = \gamma_{i}\gamma_{r}/\gamma_{tot}$ for a particular green readout power.

\pagebreak
\section*{C. Dark lifetime of NV\textsubscript{P}}
\begin{figure}[h]
\centering 
\includegraphics[width=1\linewidth]{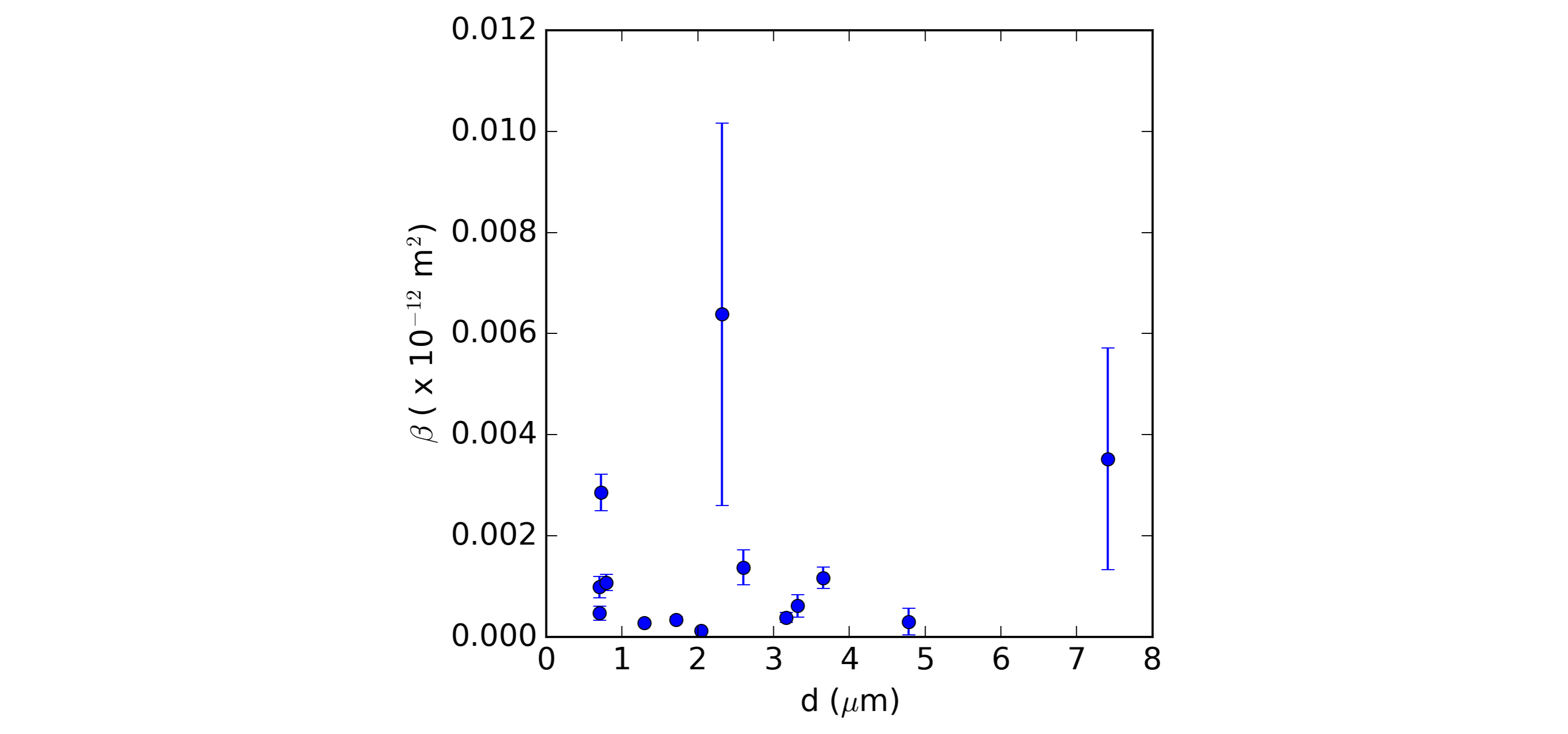}
\caption{Diagnosis of dark ionization rate of NV\textsubscript{P}. $\beta$ for 14 different NV pairs are plotted against $d$ to understand the influence of NV\textsubscript{P} dark lifetime in our experiment.
\label{fig:si_fig3}}
\end{figure}

The dark lifetime of NV\textsubscript{P} could be related to laser leakage, thermal activation, and/or tunneling of the NV center electron to the surrounding trap states (e.g. vacancy complexes, surface traps). To quantitatively understand from our obtained data points whether we can observe the dark lifetime of NV\textsubscript{P} in our experiment, we plot $\beta$ versus $d$, where $\beta = (\gamma_h/\gamma_c)2\pi d^2$ (Fig. \ref{fig:si_fig3}). From the plot, we see no gradual rise in the calculated values for longer $d$. If there was a gradual increase of $\beta$ with $d$, the increase would be the result of the dark lifetime of NV\textsubscript{P}, as at longer distances, the dark lifetime of NV\textsubscript{P} would play the dominant role in its ionization. We also do not observe any decay of $\beta$ with $d$, which again supports our conclusion of very long $L_h$.

\pagebreak
\section*{D. Verification of long diffusion length from reduced $\chi^2$}
\begin{figure}[h]
\centering 
\includegraphics[width=1\linewidth]{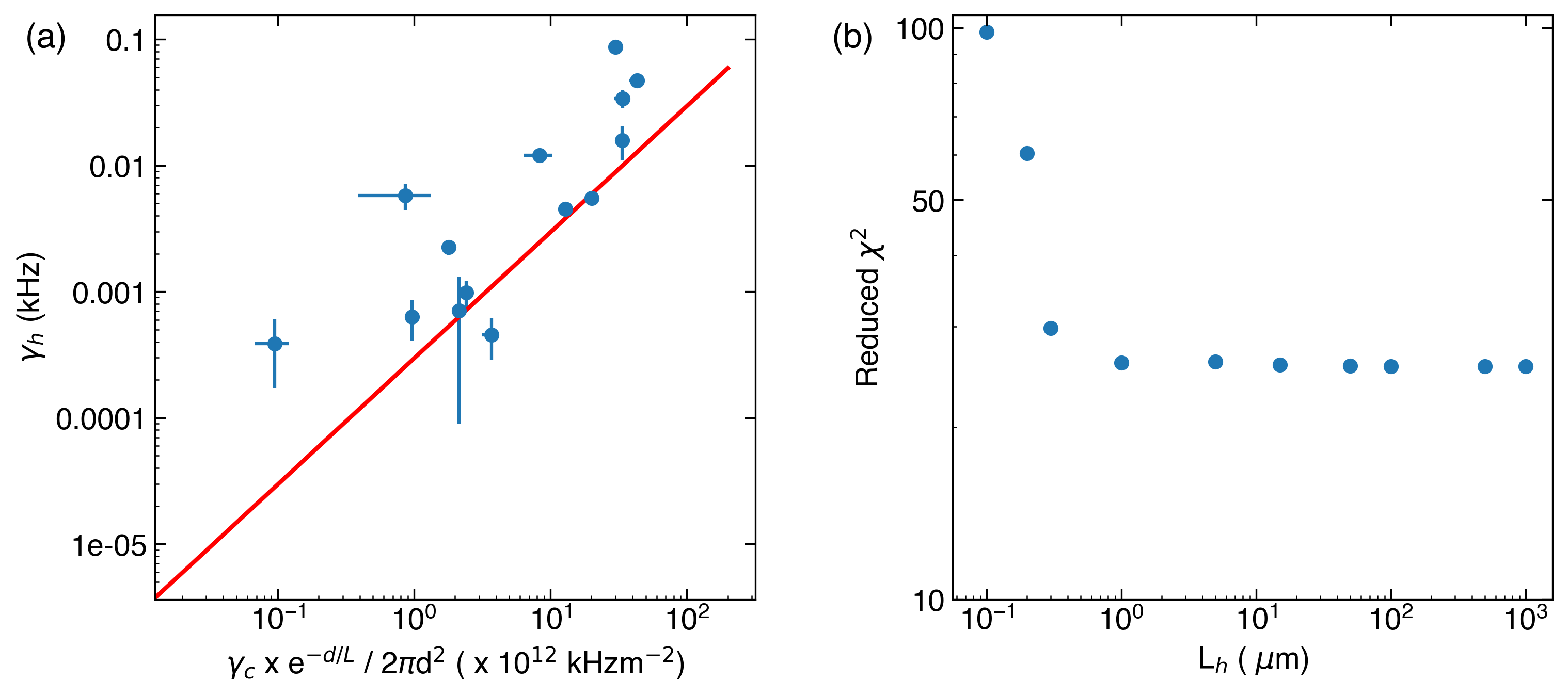}
\caption{(a) Plot used to calculate reduced $\chi^2$ for different fit with $L_h$. In this figure, we assumed $L_h$ to be 50 $\mu$m. Red solid line is a linear fit. (b) Reduced $\chi^2$ calculation from fits assuming different $L_h$.}
\label{fig:SI_fig4}
\end{figure}

To verify long diffusion length from reduced $\chi^2$, we plot $\gamma_h$ versus $\gamma_ce^{-d/L_{h}}/(2\pi d^2)$ after assuming an arbitrary value for $L_h$ (Fig. \ref{fig:SI_fig4}(a)). We then fit the calculated values with a linear fit (red solid line). From the fit we extract reduced $\chi^2$. We plot the reduced $\chi^2$ for the corresponding $L_h$ (Fig. \ref{fig:SI_fig4}(b)). The reduced $\chi^2$ reduces until it becomes saturated with increasing $L_h$. The reduced $\chi^2$ saturates for $L_h$ of $\sim$ 50 $\mu$m, which indicates that the diffusion length is much longer than the inter-NV separations we consider here.

\pagebreak

\section*{E. Tables for the rates in main text figures}
\textbf{I. Fitted values for $\gamma_h$ in Fig. 2(c) of main text:}\\

\begin{tabularx}{\textwidth} { 
  | >{\centering\arraybackslash}X 
  | >{\centering\arraybackslash}X 
  | >{\centering\arraybackslash}X | }
 \hline
 NV\textsubscript{S} excitation power ($\mu$W) & $\gamma_h$ (kHz) \\
 \hline
 138   & 0.0043$\pm$0.0001    \\
 \hline
 168   & 0.0079$\pm$0.0003    \\
 \hline
 216  & 0.0123$\pm$0.0007 \\
\hline
\end{tabularx}\\
\\
\newline  
\textbf{II. Fitted values for $\gamma_h$ in Fig. 3(b) of main text:}
\\

\begin{tabularx}{\textwidth} { 
  | >{\centering\arraybackslash}X 
  | >{\centering\arraybackslash}X 
  | >{\centering\arraybackslash}X | }
 \hline
 $d$ ($\mu$m) & $\gamma_h$ (kHz) \\
 \hline
 1.71   & 0.0049$\pm$0.0005  \\
 \hline
 1.29   & 0.0067$\pm$0.0001   \\
 \hline
 0.7  & 0.0175$\pm$0.0020 \\
\hline
\end{tabularx}